\documentclass[fleqn,12pt,twoside,usenames]{article}
\usepackage{espcrc1}
\usepackage{pst-all}
\usepackage{pst-blur}

\usepackage{graphicx}

\newcommand{\myshadowbox}{\psblurbox}

\newcommand{\nn}{\nonumber}

\newcommand{\open}{{<\kern -0.3 em{\scriptscriptstyle )}}}
\newcommand{\tv}[1]{{\bf #1}_{\scriptscriptstyle T}}

\renewcommand{\d}{{\rm d}}

\newcommand{\g}{\gamma}
\newcommand{\sig}{\sigma}

\newcommand{\Pslash}{\kern 0.2 em P\kern -0.6em \raisebox{1pt}{/}}
\newcommand{\pslash}{\kern 0.2 em p\kern -0.4em /}
\newcommand{\kslash}{\kern 0.2 em k\kern -0.5em /}
\newcommand{\Sslash}{\kern 0.2 em S\kern -0.6em /}

\def\ii{{\rm i}}

 at .4cm

\definecolor{LightGray}{rgb}{0.9,0.9,0.9}
\definecolor{LightYellow}{rgb}{0.97,0.97,0.97}
\definecolor{LightRed}{rgb}{0.9,0.9,0.9}
\definecolor{MidRed}{rgb}{1,0.6,0.6}
\definecolor{LightGreen}{rgb}{0.86,0.86,0.86}
\definecolor{MidGreen}{rgb}{0.6,1,0.6}
\definecolor{DarkGreen}{rgb}{0,0.7,0}
\definecolor{LightBlue}{rgb}{0.5,0.5,0.5}
\definecolor{MidBlue}{rgb}{0.6,0.6,1}
\definecolor{DarkBlue}{rgb}{0,0,0.7}
\definecolor{Gold}{rgb}{1.,0.84,0.}

\newcommand{\Red}[1]{{\color{Black}{#1}}}

\newcommand{\DarkGreen}[1]{{\color{Black}{#1}}}

\newcommand{\Blue}[1]{{\color{Black}{#1}}}

\psset{hatchsep=6pt,hatchwidth=.6pt}

\newpsobject{hadron}{pscircle}{unit=1truemm,linewidth=.3\psunit,
             shadow=true,blur=true}
\newpsobject{hadpolarrow}{psline}{unit=1truemm,linewidth=0.8\psunit,
             arrowsize=0pt 4.6,arrowlength=2,
             arrowinset=0,linecolor=LightBlue}

\newpsobject{quark}{pscircle}{unit=1truemm,fillstyle=solid,fillcolor=black}
\newpsobject{quarkpolarrow}{psline}{unit=1truemm,linewidth=.5\psunit,
             arrowsize=0pt 4,arrowlength=2,arrowinset=0,
             linecolor=black}

\newpsobject{fragment}{psline}{unit=1truemm,linewidth=0.4\psunit,
            linecolor=gray}

\newsavebox{\polquarkbox}
\sbox{\polquarkbox}{%
   \quark(0,0){1}
   \quarkpolarrow{->}(0,0)(7,0)
}%
\newcommand{\polquark}{\usebox{\polquarkbox}}

\newsavebox{\longpolhadronbox}
\sbox{\longpolhadronbox}{
   \hadron[fillstyle=solid,fillcolor=LightRed](-0.7,0){5}
   \hadpolarrow{->}(4.3,0)(14.3,0)
}%

\newsavebox{\transpolhadronbox}
\sbox{\transpolhadronbox}{\psset{unit=1truemm}
    \hadron[fillstyle=solid,fillcolor=LightRed](0,0){5}
    \hadpolarrow{->}(0,5)(0,15)
}%

\newsavebox{\glongpolhadronbox}
\sbox{\glongpolhadronbox}{
   \hadron[fillstyle=vlines](-0.7,0){5}
   \hadpolarrow{->}(4.3,0)(14.3,0)
}%

\newsavebox{\gtranspolhadronbox}
\sbox{\gtranspolhadronbox}{\psset{unit=1truemm}
    \hadron[fillstyle=vlines](0,0){5}
    \hadpolarrow{->}(0,5)(0,15)
}%

\newsavebox{\btranspolhadronbox}
\sbox{\btranspolhadronbox}{\psset{unit=1truemm}
    \hadron[fillstyle=crosshatch](0,0){5}
    \hadpolarrow{->}(0,5)(0,15)
}%

\psset{unit=1truemm}

\title{Spin-dependent Fragmentation Functions}

\author{Rainer Jakob\address{Universit\"at Wuppertal, Fachbereich Physik,
        42097 Wuppertal, Germany}}%
       
\begin{document}

\maketitle

\begin{abstract}
I will give an overview on fragmentation functions with particular emphasis on
spin-dependence. A straightforward classification scheme permits to label all
independent fragmentation functions for a given physical situation in an 
unambiguous way. In the context of light-cone quantisation the 
leading twist fragmentation functions have an intuitive probabilistic 
interpretation. 
\end{abstract}

\section{Motivation and Purpose of the Talk}

A confusing variety of new and exotic fragmentation functions (FFs) is 
discussed
recently, in particular in the context of the extraction of the third, leading
twist -- hitherto unknown -- nucleon distribution function. This 
{\em transversity distribution} $h_1(x)$ (frequently also 
denoted $\delta q(x)$ or $\Delta_{\scriptscriptstyle T} q(x)$) involves 
a flip of quark chirality, and 
therefore, requires a second chiral-odd function to form a chiral-even 
observable. There is a number of FFs describing 
specific physical situations which offer themselves as possible chiral-odd 
partners for $h_1(x)$.
These FFs are not only indispensable tools for the 
extraction of the {\em transversity distribution}, but of interest 
in themselves, since they
bear witness to the process of hadronisation, or in other words, how 
confinement comes about. This holds true also for the chiral-even FFs. 

The present contribution is intended to serve as a reminder that there is a
systematic and unambiguous way to classify the FFs 
according to the physical situations they describe, and to exclude those
not in accordance with general constraints from hermiticity of Dirac fields 
and their well-known behaviour under the parity transformation. In this
contribution I will restrict myself to the consideration of jets initiated by 
quarks.\\

{\em Disclaimer:} Most of the content of this contribution is based on
work done by others. I am indebted to all colleagues whose work guided my
understanding of the topic. A complete list of references would exceed the 
allowed 6 pages easily. Therefore, the references listed are to be understood 
as proposed starting points for further reading; more references can be 
found in the cited ones. 
A more comprehensive source of information is the 
WWW database~\cite{FFdatabase},  a project of the 
TMR network ESOP dedicated to {\em fragmentation functions}.

\section{Classification Scheme}
\label{sec:method}

The basic message first:\\[-3.8ex] 
\begin{itemize} 
\item
the number of independent FFs describing a certain physical situation 
is limited\\[-3ex]
\item
actually, to just a few of them at leading twist\\[-3ex]
\item
and there is a simple systematics behind.\\[-3ex]
\end{itemize} 
Let us take a look at the formal definition of FFs from soft hadronic 
matrix elements of quark field operators~\cite{Collins:1982uw}.
In a first step one defines a
quark-quark correlation function for the fragmentation process
of a quark to one (or two) hadron(s): $q\to h_1 (h_2) X$ 
\begin{minipage}{.3\textwidth} 
\begin{center} 
\scalebox{0.5}{
\pspicture(0,0)(70,60)
\psline[linestyle=dotted,dotsep=1pt](35,5)(35,40)
\pspolygon[linearc=1mm,fillstyle=solid,fillcolor=LightGreen,
           shadow=true,blur=true]
          (10,20)(60,20)(60,30)(10,30)
\rput{0}(35,25){\Large $\Delta$}
\psline[linecolor=black]{-}(16,5)(16,20)
\psline[linecolor=black]{->}(16,5)(16,15)
\psline[linecolor=black]{-}(54,20)(54,5)
\psline[linecolor=black]{->}(54,20)(54,10)
\psline[linewidth=.5pt]{->}(12,7)(12,15)
\rput{0}(7,11){$k$}
\psline[linewidth=.5pt]{->}(58,15)(58,7)
\rput{0}(62,11){$k$}
\psset{linestyle=dashed,dash=3pt 2pt,linewidth=2pt}
\psline(14,30)(21.5,45)
\psline(56,30)(48.5,45)
\rput{0}(16.5,49){$P_{1}$}
\rput{0}(53.5,49){$P_{1}$}
{%
\psset{linecolor=LightBlue}
\psline(25,30)(32.5,45)
\psline(45,30)(37.5,45)
\rput{0}(29,50){\DarkGreen{$P_{2}$}}
\rput{0}(41,50){\DarkGreen{$P_{2}$}}
}%
\endpspicture
} 
\end{center}
\end{minipage} 
\begin{minipage}{0.7\textwidth}
\begin{eqnarray} 
\lefteqn{
\Delta_{ij}(k,P_{1},\DarkGreen{P_{2}})
=\sum_X\int\frac{d^4\xi}{(2\pi)^4}\;
e^{ik\cdot\xi}\;}
\nn\\ &&\times
\langle 0|{\cal U}(0,\xi)
\Red{\psi_i(\xi)}|P_{1},(P_{2})
;X\rangle
\langle P_1,(P_{2});X|\Red{\overline\psi_j(0)}|0\rangle
\end{eqnarray} 
\end{minipage}  

Some properties of the correlation function can be derived easily, or are even
evident from its definition. The quantity $\Delta$ is a $4\times4$ matrix in 
Dirac space, and depends on the momentum vectors of the quark $k$, 
and of the observed hadron(s) $P_{1}$, ($P_2$) and possibly 
spin vector(s) $S_1$ ($S_2$) for spin-1/2 hadrons, or spin vector and 
tensor  $S_1,T_1$ ($S_2,T_2$) for spin-1 hadrons, respectively. For 
instance, for the case a single spin-1/2 hadron with momentum $P_h$ and 
spin vector $S_h$ is observed the most general ansatz can be shown to have 
the form~\cite{Ralston:1979ys,Mulders:1996dh}
\begin{eqnarray}
\label{eq:ansatz}
\lefteqn{
\Delta(k,P_h)=
\Blue{B_1} M_h + \Blue{B_2} \Pslash_h +\Blue{B_3} \kslash 
+ (\Red{B_4}/M_h)\; \sigma_{\mu\nu}P_h^\mu k^\nu }
\nn\\ &&
{}+ i\; \Red{B_5} (k\cdot\DarkGreen{S_h})\gamma_5
+ \Blue{B_6} M_h \DarkGreen{\Sslash} \gamma_5 
+ (\Blue{B_7}/M_h)(k\cdot\DarkGreen{S_h})\Pslash_h \gamma5
\nn\\ &&
{}+ (\Blue{B_8}/M_h)(k\cdot\DarkGreen{S_h})\kslash \gamma_5 
+ i\; \Blue{B_9}\;\sigma_{\mu\nu}\gamma_5 \DarkGreen{S_h}^\mu P_h^\nu
\nn\\ &&
{}+ i\; \Blue{B_{10}}\; \sigma_{\mu\nu}\gamma_5 \DarkGreen{S_h}^\mu k^\nu 
+ i\; (\Blue{B_{11}}/M_h^2)(k\cdot\DarkGreen{S_h})
\sigma_{\mu\nu}\gamma_5 k^\mu P_h^\nu
\nn\\ &&
{}+(\Red{B_{12}}/M_h)\; \epsilon_{\mu\nu\rho\sigma} 
\gamma^\mu P_h^\nu k^\rho \DarkGreen{S_h}^\sigma
\end{eqnarray}
with real amplitudes $B_i(\sigma,\tau)$ depending on the invariants
$\tau=k^2$ and $\sigma=P_h\cdot k$. Only terms are kept in the ansatz
which are in accordance with general constraints derived from\\[-3ex] 
\begin{enumerate}
\item 
the \underline{\em hermiticity} of the Dirac fields\\[-3ex] 
\item
the known behaviour of Dirac fields under \underline{\em parity transformation}.\\[-3ex] 
\end{enumerate}
We note in passing that the amplitudes $B_4$, $B_5$, $B_{12}$ are
{\em (\underline{naive}) \underline{time-reversal odd}} 
(T-odd), i.e.~they would be forbidden by a
constraint  from the behaviour of Dirac fields under time-reversal, if this
operation would not transform {\em `in'}- into {\em `out'}-states and 
vice versa. The presence of final state interactions within a current jet, and
the distortion of the {\em `out'-} states from plane waves, is sufficient to 
allow for non-vanishing T-odd amplitudes.

Fragmentation functions are obtained from the correlation 
function $\Delta$ by projection with specific Dirac matrices \Red{$\Gamma$},
and integration over components of the quark momentum
\begin{equation} 
\Delta^{[\Red{\Gamma}]}(z)
\equiv \left.
\frac{1}{4z}\int \d k^+\;
\Blue{\int \d^2\tv{k}}\;
Tr\left[\Delta\Red{\Gamma}\right]
\right|_{k^-=P_h^-/z}
\end{equation} 
or for $\tv{k}$-unintegrated FFs
\begin{equation} 
\Delta^{[\Red{\Gamma}]}(z\Blue{,-z\tv{k}})
\equiv \left.
\frac{1}{4z}\int \d k^+\;
Tr\left[\Delta\Red{\Gamma}\right]
\right|_{k^-=P_h^-/z\;\Blue{;\tv{k}}}
\end{equation} 
where it turns out that the Dirac matrices 
$\Red{\Gamma}\in\{\Red{\g^-},\;\Red{\g^-\g_5},\;\Red{i\sig^{i -}\g_5}\}$
project on leading twist FFs, and the remaining independent $4\times4$ 
matrices on higher twist FFs. Here, the notion of `twist' is used in the 
sense of a {\em `working redefinition'}, or {\em `effective twist'} as 
introduced by Jaffe~\cite{Jaffe:zw}. 
\begin{figure}[t]
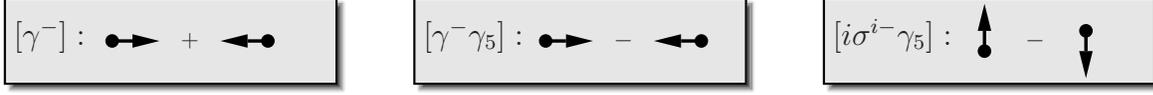
  
\begin{center} 
\begin{minipage}[t]{0.28\textwidth} 
{%
\myshadowbox[fillstyle=solid,fillcolor=LightGray]
{\begin{minipage}{.9\textwidth} 
\begin{description} 
\item[$\Red{[\g^-]:}$]
\scalebox{.45}{
\pspicture[.45](0,0)(45,20)
\rput{0}(2,10){\scalebox{2}{\polquark}}
\rput{0}(25,10){\LARGE $+$}
\rput{180}(48,10){\scalebox{2}{\polquark}}
\endpspicture
} 
\end{description} 
\end{minipage} }
}%
\end{minipage} 
\qquad
\begin{minipage}[t]{0.28\textwidth}
{%
\myshadowbox[fillstyle=solid,fillcolor=LightGray]
{\begin{minipage}{.9\textwidth} 
\begin{description} 
\item[$\Red{[\gamma^-\gamma_5]:}$]
\scalebox{.45}{%
\pspicture[.45](0,0)(45,20)
\rput{0}(2,10){\scalebox{2}{\polquark}}
\rput{0}(25,10){\LARGE $-$}
\rput{180}(48,10){\scalebox{2}{\polquark}}
\endpspicture
} 
\end{description} 
\end{minipage} }
}%
\end{minipage} 
\qquad
\begin{minipage}[t]{0.28\textwidth} 
{%
\myshadowbox[fillstyle=solid,fillcolor=LightGray]
{\begin{minipage}{.9\textwidth} 
\begin{description} 
\item[$\Red{[i\sigma^{i -}\gamma_5]:}$] 
\scalebox{.45}{%
\pspicture[.45](0,0)(40,20)
\rput{90}(5,7){\scalebox{2}{\polquark}}
\rput{0}(20,10){\LARGE $-$}
\rput{270}(35,13){\scalebox{2}{\polquark}}
\endpspicture
} 
\end{description} 
\end{minipage} }
}%
\end{minipage} 
\end{center} 
\vspace*{-6ex}
\caption{\label{fig:Gamma} The Dirac matrix $\Gamma$ determines the quark 
spin states involved in the definition of FFs from the projection
$\Delta^{[\Gamma]}$. The quark spin states projected out by the
matrices $\g^-$, $\g^-\g_5$, and ${i\sig^{\alpha-}\g_5}$ (leading twist 
projections) are schematically indicated.}
\end{figure}
An intuitive probabilistic interpretation arises in the context of light-cone 
quantisation. At leading twist the projected quark field operators can 
be rewritten as densities of ``good'' (or dynamically independent) 
components of the Dirac field operators.
Chiral projectors $P_{R/L}=(1\pm \gamma_5)/2$ allow a distinction of 
right- and left-handed components. Note that chirality and helicity are
identical for the ``good'' components of a massless quark field. 
The matrix $\Gamma$ in the projection determines quark spin 
states as sketched in Fig.~\ref{fig:Gamma}.

\section{One Unpolarised Hadron\ \protect{\cite{Ji:1993vw,Collins:1992kk}}}

\begin{figure}[hbt] 
\vspace*{-3ex}
\psframebox{
\includegraphics[width=.96\textwidth]{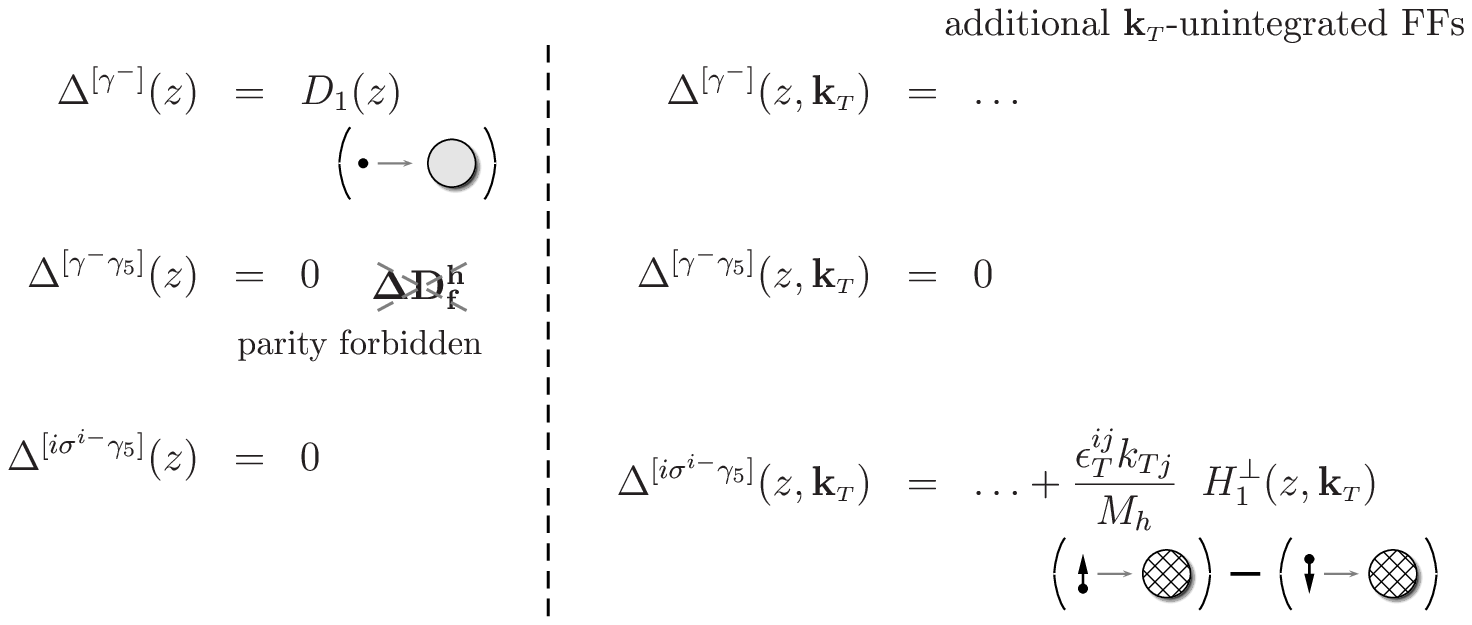}
}
\vspace*{-8mm}
\caption{\label{fig:oneunpol}Leading twist FFs for a quark fragmenting to 
one unpolarised hadron (plus rest of the jet). Integrated FFs are listed on
the left; additional unintegrated FFs on the right. Arrows indicate the
transverse spin orientation of quarks.}
\vspace*{-6mm}
\end{figure}

The independent leading twist functions describing the fragmentation
of a quark into one observed unpolarised hadron and the unobserved rest 
of the jet are listed in Fig.~\ref{fig:oneunpol}. 

There is only one FF $D_1(z)$, plus
a second unintegrated, called $H_1^\perp(z,\tv{k})$ (Collins 
function~\cite{Collins:1992kk}), occuring when transverse momentum dependent
observables are considered.  
The latter describes the correlation between the spin orientation of a
transversely polarised quark and the transverse momentum component $\tv{k}$ of
the observed hadron relative to the jet direction. 

\section{One Spin-1/2 Hadron\ \protect{
\cite{Ji:1993vw,Mulders:1996dh,Boer:1997mf,Jakob:1997wg}}}

\begin{figure}[hbt] 
\vspace*{-3ex}
\psframebox{%
\includegraphics[width=.96\textwidth]{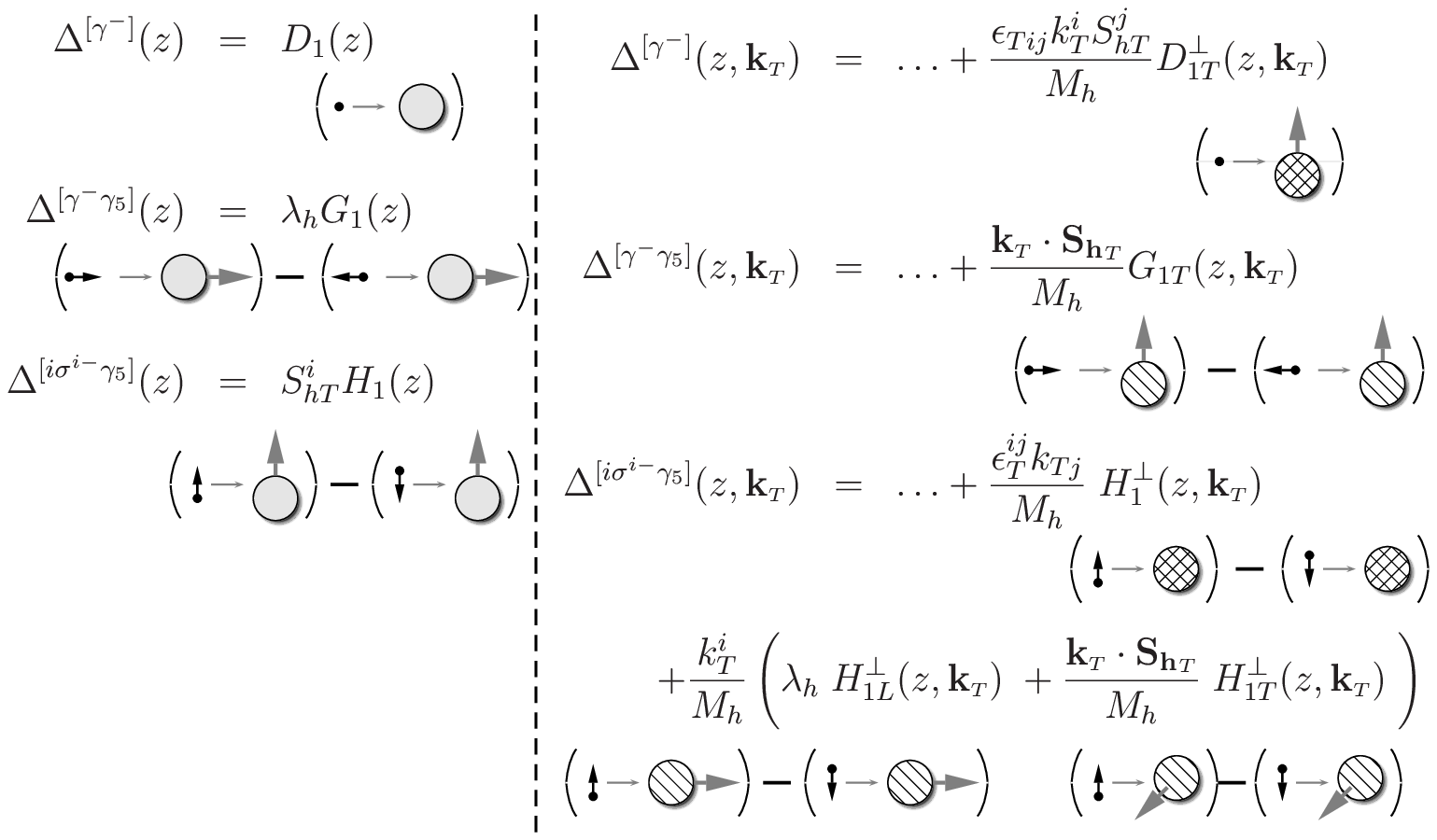}
}
\vspace*{-8mm}
\caption{\label{fig:onespinhalf} Leading twist FFs for a quark fragmenting to 
one spin-1/2 hadron (plus rest of the jet). Arrows indicate the spin
orientation of quarks and hadrons, respectively.}
\vspace*{-6mm}
\end{figure}

The independent leading twist functions describing the fragmentation
of a quark into one observed spin-1/2 hadron and the unobserved rest of 
the jet are listed in Fig.~\ref{fig:onespinhalf}.
There are two distinct groups of additional unintegrated FFs occurring in
transverse momentum dependent observables: With transverse momentum 
dependence the possibilities arise that longitudinal quark spin orientation is
correlated to transverse hadron spin $G_{1T}(z,\tv{k})$, or vice versa 
$H_{1L}^\perp(z,\tv{k})$. Or two different transverse directions, say $e_x$ 
and $e_y$ for quark and hadron spin are correlated $H_{1T}^\perp(z,\tv{k})$. 
This group is indicated by hatched hadrons in the pictograms of 
Fig.~\ref{fig:onespinhalf}. The second group (indicated with crosshatched
hadrons) comprises the functions 
$H_1^\perp(z,\tv{k})$ (Collins function) and $D_{1T}^\perp(z,\tv{k})$;
both are T-odd and can be regarded as twin partners, since they describe the
correlation of the transverse momentum of the hadron with the transverse 
quark spin, or transverse hadron spin, respectively. Both are accessible 
measuring the azimuthal dependence of the hadron production.

\section{One Spin-1 Hadron\ \protect{
\cite{Bacchetta:2000jk,Hoodbhoy:1988am,Ji:1993vw,Schafer:1999am}}}

By suitable adaption of the ansatz Eq.~(\ref{eq:ansatz}) to the case
of a spin-1 hadron the independent fragmentation functions can be derived
following the method outlined in Sec.~\ref{sec:method}. Since the spin state
of the hadron can be described by a spin vector $S$ and an additional 
tensor $T$, all FFs of the spin-1/2 case occur, plus additional ones 
associated to the spin tensor structure. The leading twist FFs are quoted in 
table~\ref{tab:spinone} (adapted from\ \cite{Bacchetta:2000jk}, where also
pictorial diagrams for a probabilistic interpretation can be found).

\renewcommand{\arraystretch}{1.2}
\begin{table}[hbt]
\vspace*{-3ex}
\caption{\label{tab:spinone}List of time-reversal even and odd, 
leading twist fragmentation functions $\Delta(\tv{k})$.}
\begin{tabular}{l@{\qquad}cc@{\qquad}cc@{\qquad}cc}
\hline\hline
 	& \multicolumn{2}{c}{$[\g^-]$}		
		&\multicolumn{2}{c}{$[\g^- \g_5]$}	&
			\multicolumn{2}{c}
{$[\ii \sig^{i-} \g_5]$}	
\\
	&T-even&T-odd	&T-even&T-odd	&T-even&T-odd	\\
\hline
U	&$D_1$	&	& 	&	& &$(H_{1}^{\perp})$ \\
L	&	&	&$G_{1L}$&	& $H_{1L}^{\perp}$&	\\
T	& &$(D_{1T}^{\perp})$&$G_{1T}$&	& \begin{tabular}{cc}
						$H_{1T}$&$ H_{1T}^{\perp}$
						\end{tabular}&	\\
LL	&$D_{1LL}$&	&	&	& &$(H_{1LL}^{\perp})$ \\
LT	&$D_{1LT}$&	& &$(G_{1LT})$	& & \begin{tabular}{cc}
	  				$(H'_{1LT}$&$H_{1LT}^{\perp})$
	     				\end{tabular} \\
TT	&$D_{1TT}$&	& &$(G_{1TT})$	& &\begin{tabular}{cc}
	  				$(H'_{1TT}$&$H_{1TT}^{\perp})$
	     				\end{tabular} \\ 
\hline\hline
\end{tabular}
\end{table}

\section{Two Unpolarised Hadrons\ \protect{
\cite{Collins:1993kq,Collins:1994ax,Artru:1995zu,Jaffe:1997hf,Bianconi:1999cd,Barone:2001sp}}}

A very promising option to access the {\em transversity distribution} 
$h_1(x)$ is to consider two hadrons produced in the same jet. There is 
one particular $\tv{k}$-integrated function, in the present scheme 
named $H_1^\open(z)$ (following~\cite{Bianconi:1999cd}) which is 
chiral-odd and  depends on  the transverse component of the
relative momentum $R=P_1-P_2$, but not on the transverse momentum 
components of the hadron pair relative to the jet direction. Thus 
a measurement of an asymmetry involving
a convolution of $h_1$ and $H_1^\open$ combines several advantages: the 
asymmetry is a
leading twist effect, collinear factorisation holds (no Sudakov
suppression), and a spin measurement of final state hadrons is not required.

\begin{figure}[hbt] 
\vspace*{-3ex}
\psframebox{
\includegraphics[width=.96\textwidth]{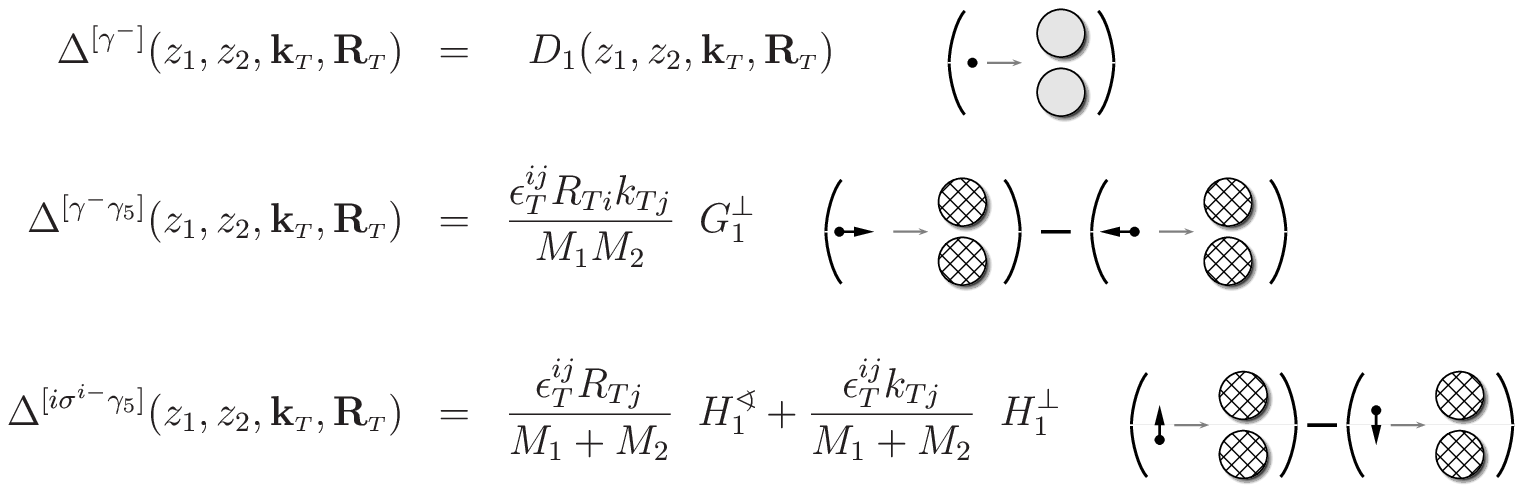}
}
\vspace*{-6mm}
\caption{\label{fig:twounpol}Leading twist FFs for a quark fragmenting to 
two unpolarised hadrons (plus rest of the jet). The functions depend on the
light-cone momentum fractions $z_1$, $z_2$ and the quantities $\tv{k}^2$,
$\tv{R}^2$, and $\tv{k}\cdot\tv{R}$, where $\tv{R}$ is the transverse 
component of $R\equiv (P_1-P_2)/2$.}
\end{figure}

Consider a situation where a quark with momentum $k$ fragments to 
two unpolarised hadrons, say a pair of pions with momenta $P_1$ and $P_2$, 
and the rest of the jet. An adaption of the method outlined in 
Sec.~\ref{sec:method} leads to the leading twist FFs depicted in 
Fig.~\ref{fig:twounpol}

\section{Generalisations and Items Not Covered}

The method outlined above lends itself to straightforward generalisations
describing other physical situations like for instance 
the observation of hadrons with spin higher than 1/2,
or the hadronisation in a gluon initiated jet~\cite{Mulders:2000sh}.  

A number of topics in the context of FFs could not be covered in the present 
short contribution, but some shall be mentioned at least (see
\cite{FFdatabase} for more details):\\[-3ex]
\begin{description} 
\item[higher twist]
The projections resulting in
higher twist FFs involve a combination of ``good'' and ``bad'' (dependent)
light-cone components of quark fields, and do not allow a simple 
interpretation. 
The ``bad'' components actually can be considered as 
quark gluon composites, which indicates the close relationship of higher twist
projections of $\Delta$ to quark-gluon-quark correlation functions.
\item[evolution]
FFs are subject to a logarithmic scale dependence similar to PDFs. Evolution 
of $\tv{k}$-unintegrated FFs has been discussed up to now only in the 
large $N_c$ limit~\cite{Henneman:2001ev}. 
\item[positivity bounds]
From the requirement of positivity of a helicity matrix a class of 
inequality relations 
between different FFs can be derived~\cite{Soffer:1994ww}. 
\item[models]
FFs have been estimated in different model calculations, among them bag
models, spectator models, and instanton models~(references to be found in 
\cite{FFdatabase}). 
\end{description}

\end{document}